\documentclass[aps,pra,10pt,twocolumn,showpacs,superscriptaddress,floatfix,footinbib,subfigure]{revtex4}
\usepackage{amsmath}
\usepackage{latexsym}
\usepackage{amssymb}
\usepackage{graphics,epstopdf,subfigure,color,hyperref}
\definecolor{Red}{rgb}{1,0,0}

\def\bra#1{\mathinner{\langle{#1}|}}
\def\ket#1{\mathinner{|{#1}\rangle}}
\def\braket#1{\mathinner{\langle{#1}\rangle}}

{\catcode`\|=\active
  \gdef\Braket#1{\begingroup
\mathcode`\|32768\let|\BraVert\left<{#1}\right>\endgroup}}
\def\BraVert{\egroup\,\mid\,\bgroup}

\def\braket#1#2{\mathinner{\langle{#1}|{#2}\rangle}}

\def\cout#1{  }

\def\R#1{{#1}}

\begin{document}

{\bf Comment on `How the result of a single coin toss can turn out to be 100 heads'} \hfill\break

 Ferrie and Combes (FC) \cite{FC} produce a classical measurement scheme that supposedly exhibits `anomalous' weak values. I   show that their model is flawed due to an  incorrect definition of the weak value. 

I begin by pointing out  that in the standard, Gaussian pointer  model \cite{AAV}, there is no need for any normalization. I then show that FC's derivation of the dichotomic classical model is flawed and leads to an incorrect normalization parameter.  Finally I show how it is possible to get a sensible result from  FC's model. This result is not anomalous. 
\vspace{6pt}

\emph{Quantum weak measurements:}  Alice prepares the state $\ket{\psi}$, gives it to Bob for a weak measurement of an operator $A$  with eigenvalues $a_k$, and then post-selects a state $\ket{\phi}$. In the standard weak measurement model (AAV)  \cite{AAV, Time} there are two free parameters:  the noise of the readout $\sigma$ and the scaling factor $x$ (see \cite{FC} eq 3) \footnote{The disturbance parameter is a function of $\sigma$ and $x$. Throughout this comment I do not refer to this parameter, however when speaking of a weak measurement I  always implicitly assume that the disturbance is small.}.  The scaling factor is  chosen in such a way that, in the strong limit $\sigma<<x$;  the pointer is sharply peaked at the eigenvalues\cite{AAV,Time,sud,AV}. If the measurement has a deterministic outcome $a_k$, Bob's pointer  will point at   the result $a_k$ with certainty (Note: if $A$ is degenerate, deterministic outcomes  do not require $\ket{\psi}$ or $\ket{\phi}$ to be eigenstates).  \R{This strong measurement limit sets the scale. It is customary to take $x=1$ so that the strong measurement results are bounded by the eigenvalues, however for any value of $x$ the  shift is bounded by the appropriately scaled  range of eigenvalues $x\cdot a_k$.   
Note: the majority of work on AAV weak measurements implicitly assume  the strong limit (and often  explicitly discuss it  e.g. \cite{AAV,Time,sud,AV,Garr,TC}). }

The AAV weak measurement  is obtained by either increasing $\sigma$ and/or  decreasing $x$ such  that for all pre and post selections in range (see below) the pointer is a Gaussian with width $\sigma$. \R{The standard choice is to fix $x=1$ and increase $\sigma$  \cite{AAV,Time,sud,AV}}. Following the logic of the strong limit  Bob can  say that the Gaussian was {\bf shifted}  by $x\cdot A_w$. \R{As in the strong limit}  only  $x$ sets the scale or normalization, if  $x$ is kept constant while $\sigma$ is changed no re-scaling  is necessary.  Note: 1)  an `anomalous' weak value is  {\bf not} the result of a normalization procedure. \R{The scale can be set at the strong limit}; 2) If $P(a_k|\phi,\psi)=1$ and we keep $x$ constant, the  meter will not shift as a function of $\sigma$.

The observed  shift   can be far outside the  range \R{measured  at the strong limit} \cite{AAV}. Experimentally, with finite  $\sigma$ , the weak value approximation will break down when $\braket\phi\psi$ is small enough \cite{limit} so  weak values are never infinite\cite{AAV,Ortho}.   An experimenter would notice deviations from the weak value approximation  by the change in the post-selected probability distribution \cite{experiment,Time}.

\vspace{5pt}

\paragraph*{The FC model}: The operational definition  of weak values \cite{AAV,Time,sud,AV,TC}  is lost  in the coarse grained model that FC consider starting  with eq 8 and ending in eq 17. The relation to a strong measurement is unclear since $\sigma$ and  $x$ are mixed into a single  `weakness parameter'  $\lambda$ that goes to $1$ for a strong measurement.

Under the right conditions,  eq 17 of \cite{FC} can be used   to {\bf calculate} the weak value from the expectation value of   $\frac{s}{\lambda}$. However, contrary to the claim of the authors eq. 17 is not ``an equivalent definition of the weak value''. 
\R{It  is easy to  provide a counterexample by modifying the superoperator in  eq. 10  so that it will produce the statistics of eq. 11  but fail to reproduce eq  17. \footnote{e.g. $\mathcal{E}_s\rho=U_s\frac{1}{2}[\rho+s\frac{\lambda}{2}(A\rho+\rho A)]U_s^\dagger$ where $U_s$ are unitary.}}

 While a coarse grained readout is  sometimes used in weak measurement experiments   (e.g. \cite{Bexp}), one starts from the definition $A_w=\frac{\bra\phi  A\ket\psi}{\braket\phi\psi}$  and normalizes the readout accordingly. FC derive their quantum model in this  way but then work in the reverse direction for the classical model,  this  makes no sense, coarse graining is not a reversible procedure. \R{Moreover the model (eq 21-23) does not even reproduce eq 10. It  is not an analogue of AAV or the coarse grained model} \footnote{In their reply \cite{rep} FC clearly state that a weak measurement followed by a conditional rotation is not the same as a weak measurement}.

\paragraph*{Can  Alice calibrate her coarse grained results?}  The classical equivalent of  $A_w$ is always in the range of eigenvalues. To have any chance of an  `anomalous value' FC must  calibrate their results according to a modified definition of the weak value. Nevertheless the fixed calibration points must at the very least be a function of both pre {\bf and} post selection, as in the quantum case. Moreover they  must  be  rooted in  the strong measurement limit.   \R{In the classical model FC set  $\lambda$ according to eq 21, which depends on pre-selection only. Moreover they  neglect to mention the strong limit.} 

 Given a set of pre and post selected  ensembles $\{\ket{\psi_i},\ket{\phi_i}\}$  the following definition can be used for a  generalized weak value.  \emph{ If a strong  measurement is expected to yield  a deterministic  result, the strong and weak value should coincide} \cite{coincide,elements} so 
\begin{equation}\label{eq}
P(a_k|\psi_i,\phi_i)=1 \Rightarrow a_w=a_k
\end{equation}
\R{This sets a  \emph{standard} range.  \emph{Anomalous} values are outside this range.}

In the  FC model the strong  limit $\lambda\rightarrow1$  implies  $P(\phi=-1|\psi=+1)=1$ and  deterministic $s=1$ so  we expect $a_w=1$ \R{and normalize accordingly}.  

\R{What surprises us about AAV is the situation where  $ |a_w|>max\{|a_k|\}$ despite the fact that  eq \eqref{eq} above holds. A classical analogue of weak values should similarly be anomalous only in cases where the result is not deterministic, i.e for all $a_k$,  $P(a_k|\psi_i,\phi_i)<1$ at the strong limit.   Since FC do not discuss this possibility, their model cannot  lead to anomalous values.}
 
\emph{Note Added:} The Reply by Ferrie and Combes contains the following
errors: 1. The condition  $x A_w \sigma <<1$  does not exist  in \cite{sud}  (where $\Delta=\frac{1}{\sigma}$) or anywhere. 2. The limit $\lambda\rightarrow 1$ implies $\delta\rightarrow 0$ \cite{FC}.

\break

\section{My response to the reply \cite{rep}}

After discarding the two technical errors (see appendix below)  there is only one substantial statement in the reply, starting with: 
\begin{quotation}
Third, Brodutch's ``counterexample" in footnote [12] is not a weak measurement of A, it is a weak measurement of A followed by a conditional rotation.
\end{quotation}
This statement is correct.   FC's  construction  (\cite{FC} eq. 21- 23)  is  also a weak measurement followed by a conditional rotation and  dephasing (see \cite{FC} eq 22). As a consequence  \cite{rep}\begin{quotation} the shift of the meter would also not correspond to the weak value. \end{quotation}   This is the point of my comment.   \hfill\break

Aharon Brodutch \hfill\break
Institute for Quantum Computing and Department of Physics and Astronomy, University of Waterloo \hfill\break

I thank the anonymous referee for helping me focus  on the normalization procedure and Eliahu Cohen, 
Asger Ipsen, Rolf Horn and Yuval Sanders for helpful  discussion and suggestions. This research was supported by  NSERC, Industry Canada and CIFAR.
\appendix

\section{Technical errors in the reply \cite{rep}}
There are two technical errors in the reply \cite{rep}. 

\subsection{The condition $x A_w \sigma <<1$   does  not exist}
FC claim that the condition   $x A_w \sigma <<1$   \emph{`has been known since the early days of weak values'}  and cite \cite{sud}. In response to my note (see  \emph{Note added} above)  they refer the reader to eq 2.45 of  \cite{kofman}. Neither of these references contains the claimed condition.  In reality  the  references cited by FC \cite{sud,kofman} give the following conditions, $\frac{A_w}{\sigma}<<1$ and  $\frac{|x A_w|}{\sigma}<<1$ respectively (for details see below).  As a consequence, FC's  claim that `\emph{ $xA_w$ cannot be an anomalous shift alone and must be re-scaled.} \cite{rep}  is unjustified. In fact this claim is simply false.  $x A_w$ can be outside the \emph{standard range}  as noted in \cite{sud,AAV} (and subsequent papers) and experimentally verified  \cite{experiment}. 

\subsubsection{Notation in \cite{sud}}

The  notation used by \cite{sud} is different from the notation   in  FC's paper (and this comment). The pointer variable is $p$, the conjugate momentum is $q$  and $\Delta$ is the spread in $q$. Consequently  the weak limit is $\Delta\rightarrow 0$ (see below eq 7 in \cite{sud}).  Converting between notations we get $\Delta=\frac{1}{\sigma}$, and \cite{sud} eq. 20 reads $\frac{A_w}{\sigma}<<1$. 

\subsubsection{Notation in \cite{kofman}}
In their subsequent  note FC reply \emph{`With regards to Brodutch’s added note, we 
refer the reader to Eq. (2.45) of Ref. [9].'} (here Ref \cite{kofman}). The referenced equation reads $|\gamma A_w|\Delta p<<1$. Again the notation is different, $q$ is the pointer, $p$ is the conjugate momentum (see table 1 and table 2 of  \cite{kofman} for notation) , $\Delta p$ is again the spread in momentum so  $\Delta p\propto\frac{1}{\sigma}$.  In our notation the condition reads $\frac{|x A_w|}{\sigma}<<1$.

\subsection{ The limit $\lambda\rightarrow 1$ implies $\delta\rightarrow 0$}
FC claim that my calculation $Pr(\phi-=1|\psi=+1)=1$ at $\lambda \rightarrow 1$ is incorrect. They say \emph{` However, in Eq. (29) of our Letter, we have $Pr(\phi-=1|\psi=+1)=1-\delta$, which is clearly independent of $\lambda$.' }. 

The claim that $\delta$ is independent of $\lambda$ is incorrect at the strong limit. In  their paper \cite{ FC} FC give the condition  $0\le\delta\le1-\lambda$ , (see \cite{FC} below eq 23).  At $\lambda\rightarrow1$ this gives $\delta\rightarrow 0$, subsequently $Pr(\phi-=1|\psi=+1)=1$ and $a_w=1$ in agreement with my calculation.


\begin{thebibliography}{99}

\bibitem{FC}
C. Ferrie and J. Combes,
Phys.  Rev. Lett. {\bf  113}, 120404  (2014).

\bibitem{AAV}
Y. Aharonov,  D.Z. Albert, and L. Vaidman, Phys. Rev. Lett. {\bf  60}, 1351 (1988).

\bibitem{Time}
Y. Aharonov and L. Vaidman, Lect. Notes Phys.  {\bf 734}, 399–447 (2008).

\bibitem{sud}
I. M. Duck, P. M. Stevenson, and E. C. G. Sudarshan Phys. Rev. D {\bf 40}, (1989).


\bibitem{AV}
Y. Aharonov and L. Vaidman Phys. Rev. A 41, 11 (1990).

\bibitem{TC}
B. Tamir and  E. Cohen Quanta  {\bf 2}: 7-17 (2013).

\bibitem{Garr}
J. L. Garretson, H. M. Wiseman, D. T. Pope and D. T. Pegg J. Opt. B: Quantum Semiclass. Opt.{\bf 6} S506  (2004). 




\bibitem{limit}
S. Pang, T.  Brun, S. Wu, and Z. Chen  Phys. Rev. A {\bf 90} , 012108 (2014).

\bibitem{Ortho}
S. Pang, S. Wu, and Z. Chen, Phys. Rev. A {\bf 86} , 022112 (2012).




\bibitem{experiment} N. W. M. Ritchie, J. G. Story, and Randall G. Hulet Phys. Rev. Lett. {\bf 66}, 1107 (1991).

\bibitem{Bexp}
D. Lu, A. Brodutch, J. Li, H. Li and R. Laflamme New J. Phys. {\bf 16} 053015 (2014)





\bibitem{elements}
L. Vaidman, Found. Phys. {\bf 26}, 895 (1996)



\bibitem{coincide}
 Y. Aharonov and L. Vaidman, J. Phys. A {\bf 24}, 2315 (1991). 

\bibitem{rep}
C. Ferrie and J. Combes,
Phys.  Rev. Lett. {\bf 114}, 118902 (2015)  

\bibitem{kofman}
 A. G. Kofman, S. Ashhab and F. Nori, Phys. Rep {\bf 520}, (2012).

\end{thebibliography}
\end{document}